\documentclass[11pt,twoside]{article}   
\date{February 21, 2005}

\title{Newtonian Gravity Predictions For Gravity Probe B}  

\author{Stanley L. Robertson\footnote{Dept. of Physics, Southwestern
Oklahoma State University, Weatherford, OK 73096}}

\begin{document}           

\maketitle                 

\begin{abstract}
Newtonian gravity and special relativity combine to produce a
gravitomagnetic precession of an orbiting gyroscope that is one
fourth as large as predicted by General Relativity. The geodetic
effect is the same in both cases.
\end{abstract}

\section{Introduction}
The similarities between Coulomb's law and Newton's law of
gravitation suggest that the underlying theories of
electromagnetism and gravitation might be similar in structure.
With the completion of electromagnetic theory by Maxwell, it was
apparent that modifications of Newtonian gravity were needed to
incorporate a finite speed of propagation of gravitational fields.
This became rather more urgent after the introduction of special
relativity with its limitations on Newtonian mechanics. Although
it might have been natural to seek minimalist relativistic
extensions of Newtonian gravity, this is not the way relativistic
gravity theory developed. In a bold and ingenious stroke, Einstein
placed General Relativity (GR) front and center as {\bf the}
relativistic gravity theory.

Although it is not widely appreciated, most of the weak field
tests of GR can be successfully passed by imposing minimalist
extensions on Newtonian gravity. However, the Gravity Probe B
experiments presently underway constitute new tests. (see
http://einstein.stanford.edu for weekly progress reports.) The
celebrated Lense-Thirring ``frame dragging'' of GR, which will be
measured by Gravity Probe B, has an interesting counterpart in
Newtonian gravity when it is extended for the magnetic effects
necessitated by special relativity. Rudimentary electromagnetic
theory and a simply extended Newtonian gravity theory are compared
here and the consequences examined for orbiting gyroscopes.

The following transformation equations can be found in many
places, e.g., p237 of ``Special Relativity'' by A.P. French [1].
They describe the transformation of quantities between the usual
inertial frames $S$ and $S'$; the latter moving at speed v down
the common +x axis direction.

\begin{equation}
x' = \gamma(x-vt) ~~x=\gamma(x'+vt')~~ t'=\gamma(t-vx/c^2)~~
t=\gamma(t'+vx/c^2)
\end{equation}
\begin{equation}
\gamma=1/\sqrt{1-v^2/c^2}~~~y=y'~~~z=z'
\end{equation}

Transformation of velocity components:
\begin{equation}
u_{x'} = \frac{u_x-v}{1-vu_x/c^2}~~~~
u_x=\frac{u_{x'}+v}{1+vu'_x/c^2}
\end{equation}
\begin{equation}
u_{y'} = \frac{u_y/\gamma}{1-vu_x/c^2}~~~~
u_y=\frac{u_{y'}/\gamma}{1+vu'_x/c^2}
\end{equation}
\begin{equation}
u_{z'} = \frac{u_z/\gamma}{1-vu_x/c^2}~~~~
u_z=\frac{u_{z'}/\gamma}{1+vu'_x/c^2}
\end{equation}
Transformation of force components:
\begin{equation}
F'_x=\frac{F_x-(v/c^2)(\bf{F \cdot u)}}{1-vu_x/c^2}~~~
F_x=\frac{F'_x+(v/c^2)(\bf{F' \cdot u')}}{1+vu'_x/c^2}
\end{equation}
\begin{equation}
F'_y=\frac{F_y/\gamma}{1-vu_x/c^2}~~~
F_y=\frac{F'_y/\gamma}{1+vu'_x/c^2}
\end{equation}
\begin{equation}
F'_z=\frac{F_z/\gamma}{1-vu_x/c^2}~~~
F_z=\frac{F'_z/\gamma}{1+vu'_x/c^2}
\end{equation}

To describe ``magnetic'' effects originating from a source of a
force field, consider two particles. {\bf Let particle 1 be
affixed to the origin of the $S'$ system and let particle 2 be
moving at velocity {$\bf u'$} relative to particle 1.} Let the
force {$\bf F'$} be the force of interaction between particles 1
and 2 and consider it as arising from the source, 1, and acting on
2.

We can write out the components of {$\bf F$} applicable in system
$S$, starting with the right members of Eqs (6),(7) and (8) and
then substituting from Eqs. (3), (4) and (5) for all components of
${\bf u'}$ to obtain
\begin{equation}
F_x=F'_x+\gamma (v/c^2)(F'_yu_y + F'_zu_z)~~~F_y=\gamma
F'_y(1-vu_x/c^2)~~~F'_z=\gamma F'_z(1-vu_x/c^2)
\end{equation}
Lump the quantities that are independent of ${\bf u}$ into a
``static'' force, ${\bf F_{st}}$ such that
\begin{equation}
{\bf F_{st}} = F'_x {\bf i}+\gamma F'_y{\bf j} +\gamma F'_z{\bf k}
\end{equation}
Then define a ``magnetic'' field quantity, ${\bf K_f}$, as
\begin{equation}
{\bf K_f} = {\bf v \times F_{st}}/c^2
\end{equation}
With these definitions, it can be easily verified that the
remaining velocity dependent terms of Eq (9) can be expressed as
${\bf u \times K_f}$. Thus we can write the force ${\bf F}$ acting
on particle 2, as observed in S as\footnote{We could make another
choice here to define ${\bf K_f}$ as a quantity having true units
of force by defining ${\bf K_f}={\bf v \times F_{st}}/c)$ and then
having ${\bf u \times K_f}/c$ as the last member of Eq. (12). This
is the choice made in gaussian units.}
\begin{equation}
{\bf F= F_{st}} + {\bf u \times K_f}
\end{equation}
\bigskip

At the instant t=0, let ${\bf F'}$ be a coulombs law force between
charges 1 and 2. Then in Eq. (10) $F'_x=kq_1q_2x'/r'^3$,
$F'_y=kq_1q_2y'/r'^3$ and $F'_z=kq_1q_2z'/r'^3$ where $k=9\times
10^9$ in MKS units. But at t=0, $x'=\gamma x$. With $y'=y$ and
$z'=z$, all components of Eq (10) have a multiplier of $\gamma$
and so ${\bf F_{st}}= \gamma kq_1q_2 {\bf r}/r'^3$ and
\begin{equation}
{\bf K_f = v \times F_{st}}/c^2 = q_2{\bf v \times E}/c^2
\end{equation}
where
\begin{equation}
{\bf E} = {\bf F_{st}}/ q_2 = \gamma k q_1{\bf r}/r'^3
\end{equation}
It should be obvious here that ${\bf B = K_f}/q_2$ is the magnetic
field at ${\bf r}$ from a charge moving with speed v in the +x
axis direction.
\bigskip

This little exercise has so far been a way of getting the usual
magnetic effects if we consider forces between moving charges, but
it is not restricted to electric forces. We are free to consider
${\bf F'}$ to be any kind of force whatever for which we can
consider particle 1 to be a source at rest at the origin of $S'$.
It could be a Newton's law gravitational force or a Hooke's law
force or a Morse potential force, or whatever. It also is not
restricted to being a central force. A dipole-dipole force with
orientation dependence could even be considered, though it would
be extremely messy. The point is that ``magnetic'' forces are
straightforwardly imposed by special relativity alone!

An important application of Eqs. (13) and (14), is that we get the
Biot-Savart law from them. All we have to do is appeal to vector
superposition. Consider a (generally curved and closed) line of
moving charge. Let $dq_1$ be an increment of charge on the line
and $d{\bf l}$ an increment of length along the line in the S
system. Since charge is invariant with respect to velocity, we can
write $dq_1{\bf v}$ as $(dq_1/dt)d{\bf l}= id{\bf l}$ and
immediately see that
\begin{equation}
d{\bf B} = (\gamma ki/c^2) d{\bf l \times r}/r'^3
\end{equation}
This is the contribution to {\bf B} at {\bf r} from a charge
moving at speed {\bf v} at $t=0$. Now it is true that if the line
of current is not straight, we need multiple frames $S'$ with
different orientations in order to calculate each contribution
d{\bf B}. But this is of minor consequence. All of the various
corresponding S frames are at rest with respect to each other. All
we have to do here is believe that we can do a vector sum of all
of the d{\bf B}'s.\footnote{In the case of a straight line of
current, the factors of $\gamma$ disappear from the integrated
{\bf E} and {\bf B} fields.} Two other things are of interest
here. First, at $t=0$ in each comoving $S'$ frame, $x'=\gamma x$
and $y'=y$ and $z'=z$. Then we have $r'=\sqrt{(\gamma
x)^2+y^2+z^2}$. In the S frame chosen for the sum of all of the
vector increments, the coordinates of the field point , 2, are
(x,y,z). Second, for ordinary currents, the drift velocity is so
small that $\gamma = 1$ to about 11 digits.

\section{Extensions to Gravity:} There is absolutely nothing to keep
us from using the gravitational force between masses $m_1$
(source) and $m_2$ (test particle) to calculate a {\bf g} field
given by
\begin{equation}
{\bf g} = {\bf F_{st}} /m_2= -\gamma G m_1{\bf r}/r'^3
\end{equation}
And again, we take the case $t=0$. We note for future reference
that $m_1$ is a rest gravitational mass in the moving $S'$ frame
and $m_2$ is the gravitational mass of the test particle as
observed in this moving frame. These distinctions hardly matter in
the circumstances considered here as we will take $\gamma =1$.

We can also define a gravitomagnetic field ${\bf B_g=K_f}/m_2$ and
have a gravitomagnetic field vector
\begin{equation}
{\bf B_g}= {\bf v \times g}/c^2
\end{equation}
And then the complete force on $m_2$ is given by
\begin{equation}
{\bf F}=m_2{\bf g}+ m_2{\bf u \times B_g}
\end{equation}
Here $m_2$ is the gravitational mass of particle 2. By considering
the gravitational forces given by Eq (7) as observed in the $S$
and $S'$ frames, one finds that these force transformation
equations require that $m_2$ must be the rest mass of the particle
even though it moves relative to the $S$ frame.\footnote{Assuming
$m_1$ to be so large that it is effectively affixed to the origin
of $S'$, consider the time interval required for $m_2$ to fall
some small way down the y axis toward the origin. Integrate
$a_y=udu/dy$ to obtain $u(y)$ and calculate the time interval as
$\int{dy/u}$. The time interval will not transform correctly
unless only rest masses are used for gravitational mass, even
though both $m_1$ and $m_2$ are moving relative to S.} For the
Newtonian force law to be the weak field, low speed limit for
gravitational forces and to retain complete consistency with
special relativity, we find that only rest mass can gravitate. The
inertial mass is, of course, $\gamma m$.

We also obtain the analogue of the Biot-Savart law as
\begin{equation}
d{\bf B_g} = (-\gamma Gi_m/c^2) d{\bf l \times r}/r'^3
\end{equation}
Where $i_m$ is the gravitational (rest) mass current in the moving
(not necessarily straight) line of mass. For present purposes we
can take $\gamma = 1$ and thus we could set $r=r'=\sqrt{
x^2+y^2+z^2}$. Again, we only need to appeal to vector
superposition in order to apply this to the calculation of
gravitomagnetic fields.

\section{Earth's Gravitomagnetic Field}
Eq. (19) can be used to calculate the gravitomagnetic field of
earth at the north pole. Here we imagine the field point for $m_2$
to be at the north pole. For simplicity, consider the earth to be
composed of spherical shells, not necessarily of the same density,
divide the shells into mass current rings rotating rigidly and
then do the laborious sum of the rings' contributions to obtain
\begin{equation}
{\bf B_g}=-\frac{GI{\bf \Omega}}{c^2r^3}
\end{equation}
where $I$ is the moment of inertia of earth about its rotation
axis, $r$ the earth radius and ${\bf \Omega}$ the earth rotational
angular velocity.

By the analogy with electromagnetism, we expect the
gravitomagnetic field to be dipolar. Thus we write the general
earth gravitomagnetic field, according to Newtonian gravity and
special relativity as
\begin{equation}
{\bf B_g} = -\frac{G [3{\bf (J \cdot \hat{r})\hat{r}
-J}]}{2c^2r^3}
\end{equation}
where ${\bf J}=I{\bf \Omega}$ is the angular momentum vector of
earth. This reduces to Eq (20) at the north pole. As expected, [2]
this last result is four times weaker than the gravitomagnetic
field of GR.

One calculation of immediate interest consists of using the
gravitomagnetic field at the north pole of earth to calculate the
gravitomagnetic effect on a Foucault pendulum placed there. The
result of this is that relative to the ``fixed stars'', the plane
of the pendulum's oscillation will precess at the rate of $B_g/2=
GJ/2c^2r^3$, in the direction of the earths rotation, where
$J=5.86 \times 10^{33} kg m^2/s$ is the angular momentum of earth
relative to fixed stars. (Note $B_g$ has units of reciprocal
time.) Numerically, this calculates out to 54 milliarcsec/yr,
which is, of course, one fourth as large as the 220 milliarcsec/yr
predicted by GR [3]. The important thing about this result is that
it is NOT ZERO. The factor of four arises from the curvature of
space-time of GR.

\section{Gyroscope Precession}
The results of the historic Gravity Probe B mission should be
published this year or next, yielding a first direct measurement
of the gravitomagnetic effect of the rotating earth on a gyroscope
in a polar earth orbit. Considering only the Lense-Thirring effect
for now and delaying the discussion of geodetic effects, the
torque on the gyroscope, according to GR will be
\begin{equation}
{\bf\tau} =\frac{d{\bf S}}{dt}=-\frac{1}{2}{\bf <B_g> \times S}
\end{equation}
where ${\bf S}=I{\bf \omega}$ is the spin angular momentum vector
of the gyroscope and ${\bf <B_g>}$ is the GR gravitomagnetic field
averaged over many orbits. For polar orbits, this is 4 times the
average over Eq (21), or $GJ/c^2r^3.$ Thus the GR precession rate
is in the direction of earth's rotation in the amount of
\begin{equation}
\omega_{LT} = GJ/2c^2r^3
\end{equation}
For a circular polar orbit at 650 km elevation, this yields a
precession rate of 0.041 arcsec/yr, with the gyroscope angular
momentum vector precessing in the direction of the earth's
rotation. Detailed calculations of the predictions of GR,
including corrections for the earth J2 quadrupole moment are
available [4]. The J2 correction is negligible for the
Lense-Thirring effect and only about one part in $10^3$, but
measurable, for the geodetic effect.

The gravitomagnetic effect of Newtonian gravity and special
relativity is easily calculated relative to nonrotating axes with
origin at earth's center. In the extended Newtonian theory, each
increment of mass $dm$ of the gyroscope experiences a
gravitomagnetic force of $dm{\bf U \times B_g}$, where ${\bf U=
v_{orbit}+u}$ and ${\bf u=\omega \times r'}$ is the velocity of
$dm$ relative to the spherical gyroscope's center of mass. The
torque relative to the gyroscope center of mass, and that causes
the gyroscope to precess is
\begin{equation}
{\bf \tau} = \int{dm{\bf r' \times \omega \times r' \times <B_g>}}
\end{equation}
which for a spherical gyroscope, reduces to Eq. (22) again. Using
Eq (21) to calculate $<B_g>$, This yields a precession in the
direction of the earth's rotation of
\begin{equation}
{\bf \omega_{precess}}=-(1/2)<{\bf B_g}>=G{\bf J}/8c^2r^3
\end{equation}
which is exactly one fourth of the GR result.

\section{Geodetic Precession}
Although one may think of Newtonian gravity as something existing
in flat spacetime, there is a certain sense in which space must be
curved. Light slows down in gravitational fields and meter sticks
shrink and clocks run slow in gravitational fields. In effect, the
metric of space-time is altered. Space is inherently curved in the
sense that the ratio of a circumference of a circle centered on a
gravitating mass to its measured radius will not be $2\pi$ because
the parts of the measured radius closest to the mass will be
measured with a shrunken meter stick. Interestingly, the speed of
light, measured with shrunken meter sticks and clocks running slow
will yield the free space speed at any location and life will seem
normal anywhere.

To account for the classic tests of GR, such as the perihelion
shift of planet Mercury, an extended Newtonian theory must
encompass decreases in the speed of light in gravitational fields.
This is all that is needed to account for the weak field results
that have historically been attributed to GR. But this extension
is non-trivial when coupled with special relativity, because of
the alterations of our measures of length and time. In many
theories of gravity in otherwise flat space-time (e.g Krogh's
deBroglie wave refraction theory [5], Puthoff's polarizable vacuum
theory [6], Krogdahl's flat spacetime theory [7], Rastall's
minimally extended Newtonian theory [8], the 1958 Yilmaz theory
[9], etc) the changes of measures of length and time are
equivalent to having an exponential metric of the form:
\begin{equation}
ds^2=c_o^2dt^2 e^{2 \phi}-e^{-2\phi}(dx^2+dy^2+dz^2)
\end{equation}
Here $\phi(x,y,z)$ is the gravitational potential (an
intrinsically negative quantity) divided by $c_o^2$, with $c_o$
the free space speed of light in regions free of gravitational
fields. It can be shown that particles (including photons)
subjected to only gravitational forces follow geodesics in this
metric, which thereby accounts for all of the previous weak-field
tests of General Relativity. The exponential metric depends only
on the fact that Newtonian potentials are specified only to within
an arbitrary additive constant, as shown by Rastall [8]. In
contrast, the event horizons of GR's Schwarzschild coordinates
black holes depend explicitly on an absolute potential of $\phi=
1/2$, which is without parallel in the rest of accepted physics.
The exponential metric can also be derived exactly from the
principle of equivalence of gravitational fields and accelerated
reference frames in special relativity.

In parameterized post-Newtonian (PPN) approximation, the geodetic
effect depends only on the diagonal terms of the metric tensor and
on the parameter $\gamma_1$, where the expansion of the time-time
metric coefficient yields $g_{tt}=1+2\phi+2\gamma_1 \phi^2+....$.
In the isotropic coordinates form of the metric, General
Relativity yields $\gamma_1=1$. Since the expansion of
$g_{tt}=e^{2\phi}=1+2\phi+2\phi^2+...$, it is apparent that
$\gamma_1=1$ is expected in ``flat" spacetime theories. Thus they
will predict the same geodetic effect as GR. Ciufolini and Wheeler
[2] give the geodetic gyroscopic spin precession of Gravity Probe
B as
\begin{equation}
{\bf \omega_{geo}}=-\frac{\bf V}{2c^2} {\bf \times
\frac{dV}{dt}}+(1+\gamma_1){\bf V \times \nabla}\phi
\end{equation}
where ${\bf dV/dt=a+\nabla} \phi$, with ${\bf a}$ representing any
non-gravitational accelerations and ${\bf V}$ the orbital
velocity. The first term of Eq (28) is the usual Thomas precession
of any vector taken along an accelerated (orbital) path. The
combination of first and second terms comprises the geodetic
effect. With $\gamma_1=1$ in both GR and minimally extended
Newtonian gravity there should be agreement on the geodetic
effect. The geodetic effect has been measured to about 1\%
accuracy prior to Gravity Probe B.

It is difficult to say what to expect from other tensor theories,
such as the 1971 Yilmaz theory [10]. The gravitomagnetic effects
of GR cannot be calculated from the Schwarzschild metric. A Kerr
metric is required since  the gravitomagnetic effects arise from
the off-diagonal elements of the metric tensor, of which there are
none in Schwarzschild geometry. I have no idea how to calculate
the analogue of the Kerr metric in the Yilmaz theory, but I
suspect that its off diagonal elements will agree with the Kerr
metric to first order, which is all that will be tested in the
earth's field. So it is possible that GP-B might confirm both the
Yilmaz theory and GR, but on the other hand, if a correct gravity
theory is a vector field theory in flat space-time, then GP-B
should confirm the simpleminded approach given here; i.e., results
one fourth as large as those expected from GR. Lastly, there is a
geodetic effect from the earth orbit around the sun that would
produce about 19 milliarcsec/yr precession of the Gravity Probe B
gyroscopes [3]. This is in a direction perpendicular to the
ecliptic, which puts it very nearly in the same direction as the
GR Lense Thirring effect. Thus, the Newtonian gravity and special
relativity result would combine to yield a precessional angular
velocity component along earth's axis of about $.01+.019
cos(23.5^o) \sim .0275$ arcsec/yr parallel to the earth's rotation
axis.

In the geodetic effect the Thomas Precession term contributes one
fourth of the effect of spacetime static curvature and with
opposite sign. These effects are essentially the outcomes of a
Fermi-Walker transport of a spin vector around an accelerated path
and in curved space. In the Lense-Thirring effect, there should be
no corresponding combined effects as gravitational forces are
supposed to be entirely encompassed within GR. A ``Grateful Dead
exclusion rule"\footnote{e.g., wear a shirt or a tie or neither
but not both.} should apply here. Either the GR or Newtonian
effects (or neither?) should be observed, but not a combination of
both.

\bigskip
\noindent
{\bf Acknowledgements:}

I am indebted to Ray Jones and
Charles Rogers for helpful critiques. I thank Jose Maciel Natario
for pointing out an error in the first draft of these notes.

\end{document}